%% file: NRAO150_v7.tex
\documentclass[preprint]{aa}
\usepackage{graphicx}
\usepackage{natbib}
\usepackage{txfonts}
\usepackage{color}

\bibpunct{(}{)}{;}{a}{}{,}  

\newcommand {\apgt} {\ {\raise-.5ex\hbox{$\buildrel>\over\sim$}}\ }
\newcommand {\aplt} {\ {\raise-.5ex\hbox{$\buildrel<\over\sim$}}\ }

\newcommand{\micron}{$\mu$m\ }
\newcommand {\ia}{\'\i }

\begin{document}

\title{The redshift and broad band spectral energy distribution of
\object{NRAO~150}\thanks{Based on observations made with the William Herschel Telescope 
operated on the island of La Palma by the Isaac Newton Group in the 
Spanish Observatorio del Roque de los Muchachos of the Instituto de Astrof\'{i}sica de Canarias.}
}
\author{
       J.~A. Acosta--Pulido\inst{1,2}
       \and
       I. Agudo\inst{3,4}
       \and
       R. Barrena\inst{1,2}
       \and
       C. Ramos Almeida\inst{1,2,5}
	  \and 
	  A. Manchado\inst{1,2}
	  \and 
	  P. Rodr{\'i}guez--Gil\inst{1,6}
       }

\offprints{J.~A. Acosta--Pulido, \email{jap@iac.es}}

\institute{Instituto de Astrof{\'i}sica de Canarias (IAC), 
           C/V{\'i}a L{\'a}ctea, s/n, E-38200, La Laguna, Tenerife, Spain
      \and
          Departamento de Astrof{\'i}sica, Universidad de La Laguna, E-38205 La Laguna, Tenerife, Spain
      \and	   
          Instituto de Astrof\'{\i}sica de Andaluc\'{\i}a (CSIC),  
              Apartado 3004, E-18080 Granada, Spain
      \and	   
          Institute for Astrophysical Research, Boston University, 
          725 Commonwealth Avenue, Boston, MA 02215, USA 
      \and
          Department of Physics \& Astronomy, University of Shefield, UK
      \and 
      	   Isaac Newton Group of Telescopes, La Palma, Spain	    	       
          }

\date{Received:23/12/2010 ; accepted: 09/03/2010}

\titlerunning{The redshift and SED of NRAO 150}

\abstract
{\object{NRAO~150} is one of the brightest
radio and mm AGN sources on the northern sky. It has been revealed as an interesting 
source where to study extreme relativistic jet phenomena.
However, its cosmological distance has not been reported so far, 
because of its optical faintness produced by strong Galactic extinction.}
{Aiming at measuring the redshift of \object{NRAO~150}, and hence to start making possible quantitative studies from the source.}
{We have conducted spectroscopic and photometric observations of the source in the near-IR, as well as in the optical.}
{All such observations have been successful in detecting the source. The near--IR spectroscopic observations reveal strong H$\alpha$ and H$\beta$ emission lines from which the
cosmological redshift of \object{NRAO~150} ($z=1.517\pm0.002$) has been determined for the first time.
We classify the source as a flat--spectrum radio--loud quasar, for which we estimate a large super--massive black--hole mass $\sim5\times 10^{9} \mathrm{M_\odot}$.
After extinction correction, the new near-IR and optical data have revealed a high-luminosity continuum-emission 
excess in the optical (peaking at $\sim2000$\,\AA, rest frame) that we attribute to thermal emission from the accretion disk for which we estimate a 
high accretion rate, $\sim30$\,\% of the Eddington limit.}
{Comparison of these source properties, and its broad--band spectral--energy distribution, with those of \emph{Fermi} blazars allow us to predict 
that \object{NRAO~150} is among the most powerful blazars, and hence a high luminosity --although not detected yet-- $\gamma$--ray emitter.}


\maketitle

%

\section{Introduction}
\label{Int}

\object{NRAO~150}, first catalogued by \citet[]{Pauliny66}, is nowadays one of the
strongest radio and mm AGN sources in
the northern sky  \citep[e.g.][]{Terasranta05,Agudo:subm}.
The source has been monitored at cm and mm wavelengths for decades \citep[e.g.,
][ and references therein]{Aller85,Reuter97,Terasranta04}, and has displayed
flux densities in the range $[2, 16]$\,Jy at 2\,cm\footnote{{\tt
http://www.astro.lsa.umich.edu/obs/radiotel}} and $[1.5, 9.5]$\,Jy at
3\,mm\footnote{H. Ungerechts, private communication}, with absolute maxima at
the beginning of year 2009.

At radio wavelengths, on VLBI scales, \object{NRAO~150} displays a compact core
plus a one-sided jet extending up to $r \apgt 80$~mas with a jet structural
position angle (PA) of $\sim 30^\circ$ \citep[e.g., ][]{Fey00}.
The first set of ultra-high-resolution mm-VLBI images of \object{NRAO~150}
\citep{Agudo07} have allowed 
to report a large misalignment ($>100^{\circ}$) between the cm-wave and the
mm-wave jet, which is, together with the one sidedness of the jet, a clear sign
of jet orientation close to the line of sight.
More intriguing is the evidence of fast ``jet wobbling'' at
$\sim$11$^{\circ}/$yr in the plane of the sky; the fastest reported for an AGN
so far.
The observations by \citet{Agudo07} together with the cosmological redshift
measurement presented in this paper and non-contemporaneous X-ray data have allowed
to report the first quantitative estimates of the
basic physical properties of the inner jet in \object{NRAO~150}; i.e., Doppler
factor $\delta \approx 6$, the bulk Lotentz factor $\gamma \approx 4$, the angle
subtended between the jet and the line of sight $\phi \approx 8^{\circ}$,  and
the magnetic field intensity of the flow $B\approx 0.7$\,G.
Such large $B$ estimate seems to be compatible with the highly non-balistic
superluminal motion of the inner jet in the source revealed by the mm-VLBI
images ($\beta_{\rm{app}}\approx3$ times the speed of light), which have also
shown \object{NRAO~150} as a prime target to study the origin of the jet wobbling
phenomenon \citep{AgudoVSOP}.

\object{NRAO~150} was first detected in December 1981 in the optical by
\citet{Landau83}.
However, no optical classification or distance determination has been reported
so far, perhaps due to the difficulties to observe the source in the visible
range, where \object{NRAO~150} is strongly affected by Galactic extinction
(Galactic latitude  $b\approx-1.6$\,$^{\circ}$).
This problem is partially overcome in the near-IR range, where spectroscopic
observations from strong lined objects can be performed from Earth.
Independently of the Galactic absorption along the line of sight of the source, the
near-IR range is the adequate spectral range to detect the strong H$\alpha$ line
from AGN at cosmological redshifts between 1.2 and 3.6 \citep[e.g.,
][]{Babbedge04}.

Here, we present the results of our spectroscopic near-IR observations, which were successful on detecting, 
for the first time, emission lines from \object{NRAO~150}. 
In Sect.~\ref{Obs} such observations and their data reduction procedures are
outlined, together with a set of optical and near-IR photometric observations
performed in the 2005--2007 time span.
The cosmological redshift determination, the classification, the first estimate of the mass of the super-massive 
compact object in \object{NRAO~150} and its accretion rate, as well as its broad band spectral 
energy distribution, are presented and discussed in
Sect. \ref{sc:results} and \ref{sc:disc}, whereas a summary of our main results and our conclusions are
provided in Sect.~\ref{sc:conclu}.


\section{Observations and Data Reduction}
\label{Obs}

\subsection{Near Infrared spectroscopy}

We have obtained near-IR spectroscopy of \object{NRAO~150}
using the Long-slit Intermediate Resolution Infrared Spectrograph (LIRIS,
built at the Instituto de Astrof\'{i}sica  de Canarias
\citealt{Manchado04,Acosta03}) on the 4.2~m William Herschel Telescope (WHT) in 
two epochs (2005 March and 2007 January).

\begin{figure*}
\begin{center}
\includegraphics[height=15truecm,width=8truecm,angle=90]{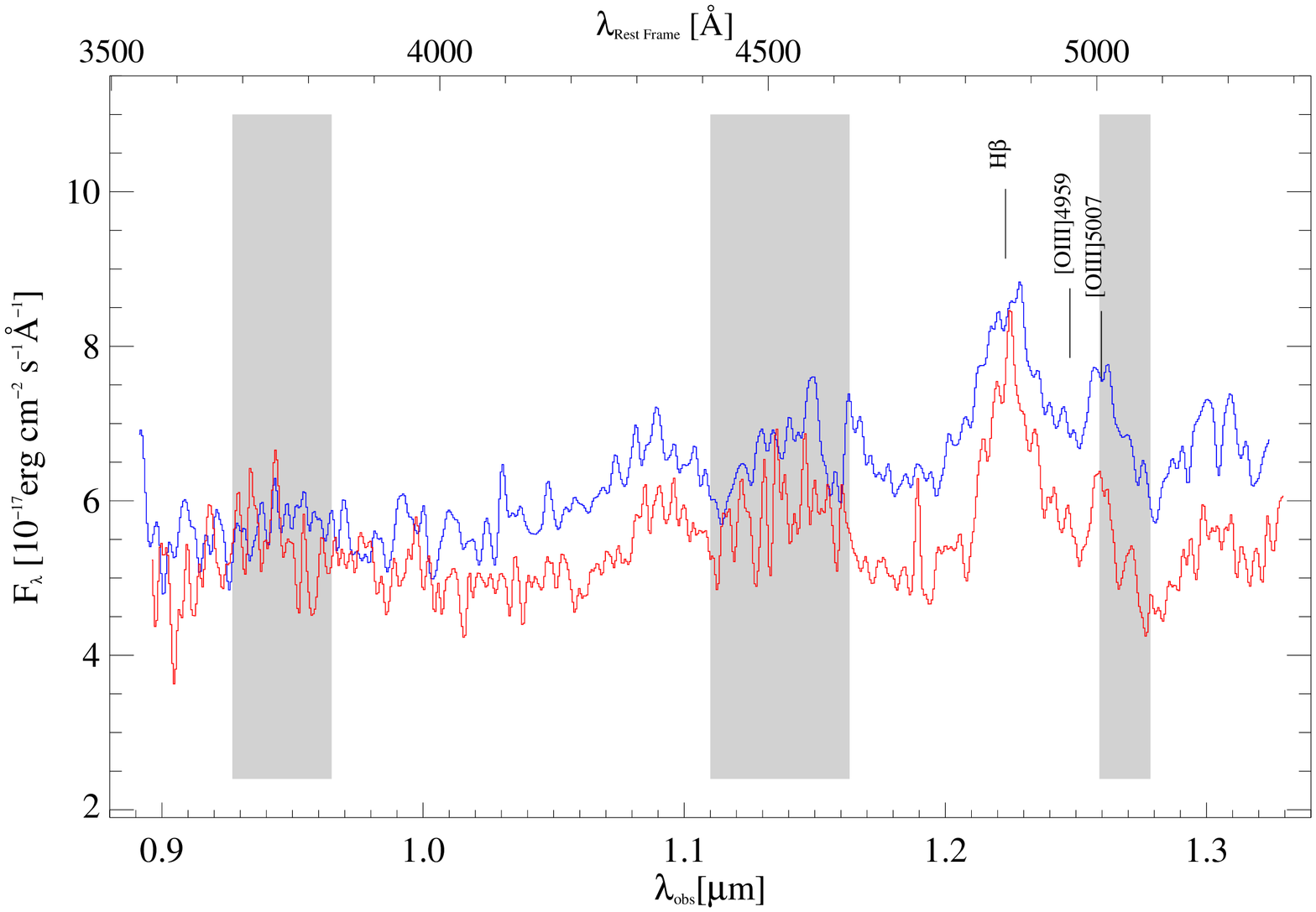}
\includegraphics[height=15truecm,width=8truecm,angle=90]{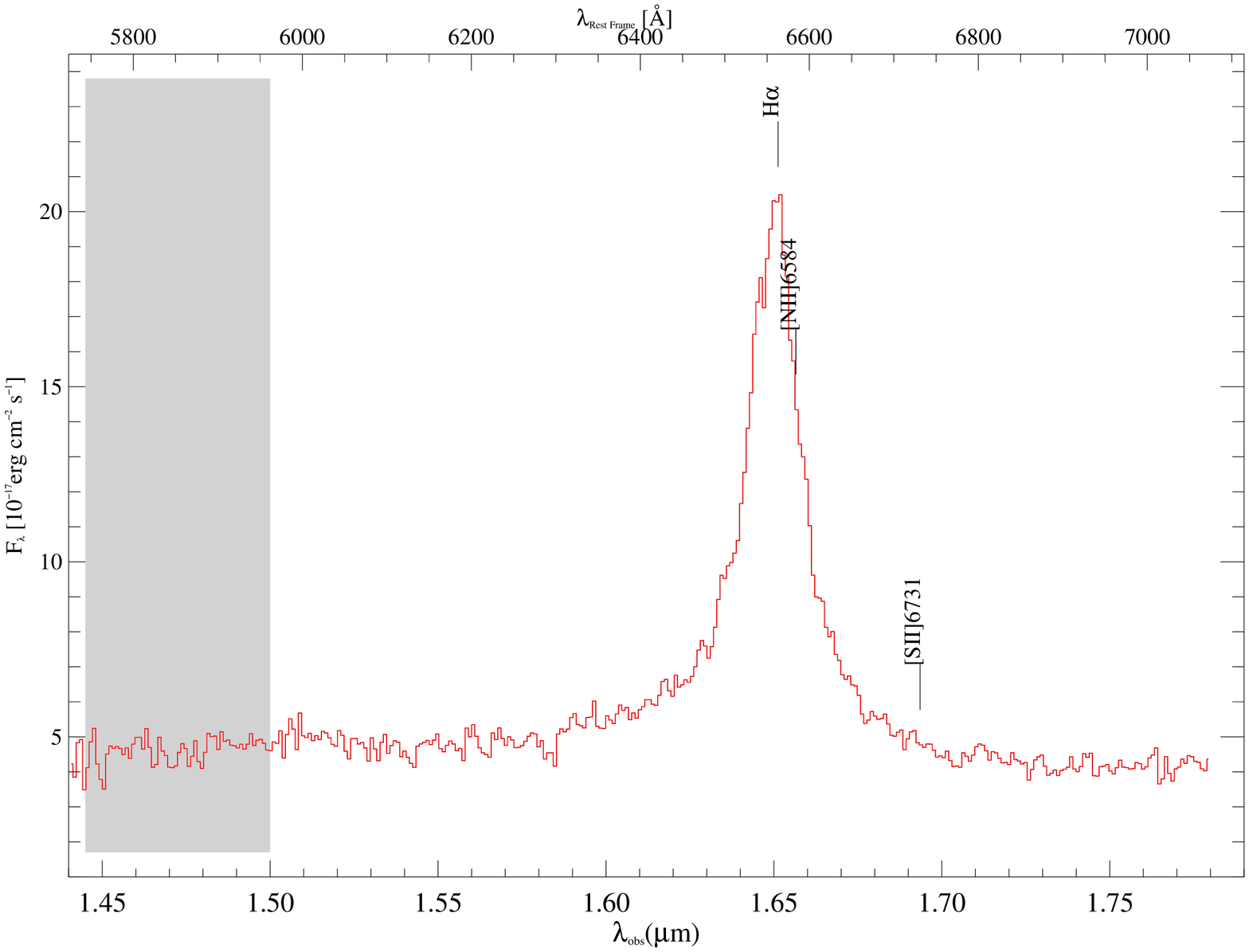}
\includegraphics[height=15truecm,width=8truecm,angle=90]{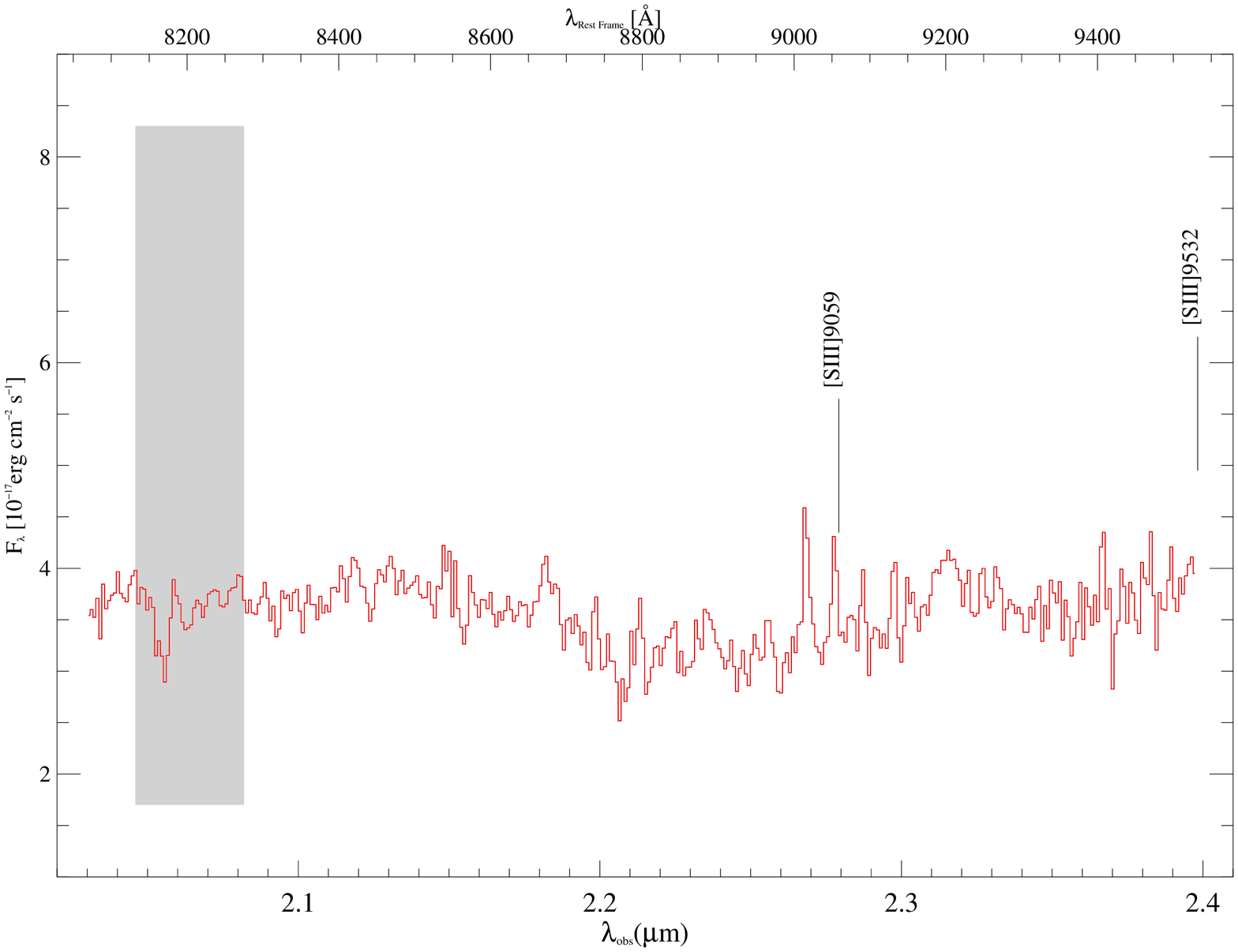}
\end{center}
\caption{\footnotesize{Final reduced spectrum of \object{NRAO~150}. 
The position of the most prominent features are marked. Deep atmospheric absorption bands are indicated by
vertical bands.
Top panel:
Spectra corresponding to the bands
Z+J observed in two epochs. The thicker red color line corresponds to 2005 Mar,
and the thinner blue color line corresponds to 2007 Jan.  
Middle and bottom panels: Spectrum corresponding to the bands H and K, respectively.} 
\label{fi:speczj_hk}}
\end{figure*}

The spectra were obtained using grisms LR\_ZJ (bands Z+J) and LR\_HK (bands H+K).
For the observations in 2005, the slit width was selected to be $0\farcs 75$, providing resolutions 
of 700 and 600 in the Z+J and H+K bands, respectively. 
For those in 2007 the slit width was 1" to match the seeing, providing a resolution of 500 in the Z+J bands.
The slit orientation ($\mathrm{PA}=118.8^\circ$) was chosen to include in the slit
aperture both NRAO~150 and the object 2MASS J03592889+5057547.  
The observations were performed following an ABBA telescope nodding pattern, 
with each AB cycle repeated 3 times. The exposure time for a single frame 
was  600~s, giving a total of 3600~s for both spectral ranges. 
The data were reduced following standard recipes for near-IR spectroscopy, using the dedicated software {\it lirisdr}, 
developed within the IRAF\footnote{IRAF (Image Reduction and Analysis Facility) is distributed by the National Optical 
Astronomy Observatories, which are operated by AURA, Inc., under cooperative agreement with the National Science Foundation.} 
enviroment by the LIRIS team.  The basic reduction steps comprise sky
subtraction, flat-fielding, wavelength calibration, and finally, the combination of
individual spectra by the common {\it shift-and-add} technique. 
For a more detailed description of the reduction process, see \citet{RamosAlmeida06,RamosAlmeida09}.
The flux calibration and telluric absorption correction are usually determined
from observations of  nearby stars of spectral types A0V or G2V, which are obtained 
inmediately after or before the science observations. 
In the first epoch the selected correction star resulted to be a multiple
stellar system whose 2MASS near-IR colors indicate a spectral type different than A0.
We have determined a spectral response function using the spectrum of the star 2MASS J03592889+5057547, 
which was observed simultaneously with \object{NRAO~150}.
The stellar spectral shape was corrected by 
a blackbody of temperature corresponding to an M5 spectral type.
A posteriori we derived the telluric correction from the normalized spectrum of 
star 2MASS J03592889+5057547. 
During the second epoch an A0V star was successfully observed.
A modified version of the Xtellcor routine was used \citep{Vacca03} to obtain the calibrated flux
and telluric corrected \object{NRAO~150} spectrum.  Relative light losses due to atmospheric
differential refraction are very small, always below few percent, despite slit was not oriented at the 
parallactic angle.
The flux calibrated spectra are presented in Fig.~\ref{fi:speczj_hk}.

\input{table1}

\subsection{Near Infrared photometry}

We obtained images in the  J and K$_s$ filters using LIRIS at the same epochs as the near-IR spectroscopy. 
In both cases the images were obtained following a 5 point dithern pattern. 
Individual frames of 20 and 10 sec were taken in the J and Ks filters, 
respectively, giving as total exposure times those reported in Table \ref{ta:phot}. 
The images were reduced using the task {\it ldedither} included in the dedicated software package {\it lirisdr}, 
developed under IRAF by the LIRIS team.  The main data reduction steps are 
sky subtraction, flat-field correction and finally a combination of the images after proper alignment. 

The photometric calibration was determined by comparison with 
the 2-MASS catalogue \citep{Cutri03}. Zero points for 
our J and K$_s$ photometry were determined using about 20 stars with known 2-MASS magnitudes (brighter than $J \sim 15.5$)
included in the LIRIS field. The typical dispersion between instrumental and 2-MASS magnitudes is $\sim 0.08$. 

\input{table2}

\subsection{Optical Photometry}

\begin{figure}
\includegraphics[width=8.7cm]{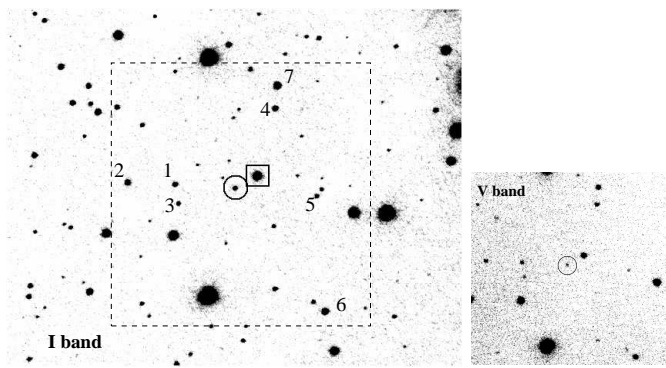}
\caption{Identification charts of \object{NRAO~150} in I and V filters.
The circle and square symbols mark \object{NRAO~150} and the star 2MASS~J03592889+505754, used as reference for spectroscopy. 
Dashed square represent the plotted area in V band. North and East are toward the top and left of the frame, respectively. Numbers 
note the calibration stars proposed for this field. \label{fi:idchart}}
\end{figure}

In addition to the near-IR observations, we also obtained images in the optical range using several
telescopes at different epochs.  The first observations were performed using the CCD camera mounted on the IAC-80 at the 
Teide Observatory during the nights of 2005 Nov 2 and 11. 
We used filters V, R and I  (see Table \ref{ta:obslog}). 
Our target was detected in all bands (see Fig. \ref{fi:idchart} for examples of I and V images). 
Given the faintness of the source, a larger diameter telescope, the 2.5~m Liverpool robotic telescope \citep{Steele04}
was used for the remaining observations. Filters Johnson V and Sloan r' and
i' were used,  although most data were obtained in r' only (see Table~\ref{ta:obslog}).

All data were reduced in IRAF following the standard procedure, i.e.,
bias subtraction, flat-fielding, and image 
combination. The photometric calibration was based on images of Landolt fields
obtained at photometric conditions. The photometry was obtained using the  SExtractor program 
\citep{Bertin96} and the AUTOMAG parameter. The resulting photometric measurements are listed in Table \ref{ta:phot}. 
We estimated that the detection of faint objects in our images is complete down 
to $V=23.1$ (24.0) mag, $r'=21.7$ (23.0) mag and $i'=20.4$ (21.5) mag for  $S/N=5$ (3) within the observed fields.

Based on our photometric calibration, we propose seven reference stars for $r'$ and $i'$ filters which can be used 
for future monitoring. These stars are marked in Figure \ref{fi:idchart} and their corresponding magnitudes are listed in Table
\ref{tab:comp}.

\input{table3}

\section{Results}
\label{sc:results}

\subsection{Near IR spectrum}

A very prominent feature is observed in the spectrum of \object{NRAO~150} in the
H band at 1.65~\micron (see Fig. \ref{fi:speczj_hk}). 
A less intense, but notable, feature is also seen in the J band at $\sim 1.23$~\micron. 
These are identified with the H$\alpha$ and H$\beta$ lines, redshifted by $z=1.517 \pm 0.002$
and then providing the first redshift determination of \object{NRAO~150}.
The corresponding luminosity distance of the source is 
$d_{\rm{L}}=11.2\times10^4$\,Mpc, for a
$H_o=71$\,km\,s$^{-1}$\,Mpc$^{-1}$, $\Omega_{m}=0.27$, and
$\Omega_{\Lambda}=0.73$ cosmology\footnote{The cosmology calculator available in the Web
({\tt http://www.astro.ucla.edu/$\sim$wright/CosmoCalc.html}) was used.}.

We measured the center, width and flux of the narrow and broad components of H$\alpha$ and H$\beta$ by fitting 
two Gaussian components plus a slope as the continuum (see Table \ref{ta:fitlines}).
We started by fitting H$\alpha$ since this part of the  spectrum has higher
signal-to-noise ratio and is not contaminated by other features such as FeII
emission. The H$\alpha$ profile is well fitted by a narrow component of
FWHM=170\AA\ (1458~km/s  rest-frame) and a broad component of FWHM=666~\AA\ (5745~km/s rest-frame).
The broad component is blueshifted with respect to the narrow component by 62\AA\ (532~km/s). 
In the case of H$\beta$ the S/N ratio is much lower and was not possible to let
vary the line widths. Instead they were constrained from the best fitting value of H$\alpha$.  

\begin{figure}
\includegraphics[width=8.7cm]{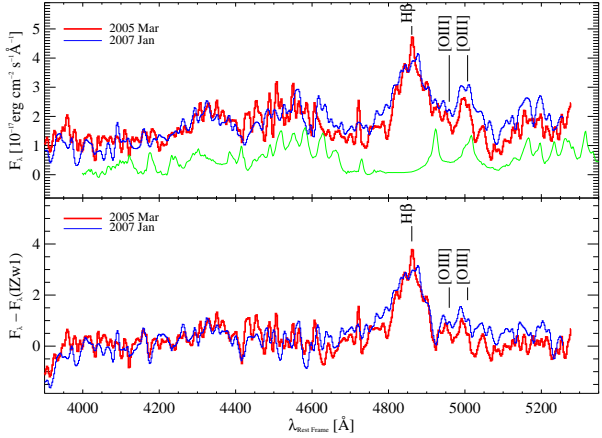}
\caption{\footnotesize{} Top panel: The two NRAO~150 spectra  (histogram style line) taken at different
epochs are represented together with a scaled version of the IZw1 template  (green continuous thin line).
Bottom panel: The spectra  after subtraction of the template. It can be seen
that the residual spectra consist basically on a broad H$\beta$ feature, a
shallow  bump at 4300\AA\ corresponding to the Balmer jump, and very weak [OIII]
lines. \label{fig:subFeII}}
\end{figure}

Other broad features showing lower S/N are observed around 4300~\AA\ (rest frame), corresponding to the Balmer limit,
and at 4450-4700, 4924, 5018 and 5150-5350~\AA, associated to blends of FeII emission (see Fig. \ref{fi:speczj_hk}).  
In order to remove and confirm the importance of FeII emission we have subtracted the empirical FeII emission template
of \citet{Boroson92}, scaled to the intensity of the observed features (see
Fig~\ref{fig:subFeII}). This template was generated from a spectrum of
PG~0050+124 (IZw~1), which is well known by the strength of its FeII emission
and the narrowness of its H recombination lines \citep{Oke79}. The template 
(kindly provided by Dr. P. Marziani) was prepared by removing the lines which are not associated with Fe~II transitions. 
It can be seen from Fig.~\ref{fig:subFeII} that after the FeII template 
subtraction the residual spectra show uniquely the broad H$\beta$ feature, plus
a shallow bump at 4300\AA\  corresponding to the Balmer limit. 
Given the radio loud AGN properties of \object{NRAO~150},
it is expected to observe intense narrow forbidden emission lines such 
as [OIII]~$\lambda$~5007\AA, [NII]~$\lambda$~6548\AA, $\lambda$~6584\AA, and 
[SII]~$\lambda\lambda$~6716, 6731\AA.
However, none of these features are detected at a significant level over the continuum. 
The implications of weak or absent typical NLR features will be discussed below in section \ref{sc:weakNLR}.

\subsection{Optical and near IR variability}
\label{sec:varia}
Our optical and near-IR photometric measurements span a range of almost 2 years (see Table~\ref{ta:phot}),
which is typically a sufficiently long time range to detect intrinsic variability in radio loud AGN
\citep[e.g.,][]{Terasranta04,Terasranta05}. Indeed, \object{NRAO~150} shows a nearly monothonic radio 
flux increase by a factor $\sim4$ from 1991 to 2005 \citep{Terasranta04,Terasranta05}.   
However, we cannot  make any firm claim about optical variability based on measurements in the r' band, by far the band 
with the longer and denser time coverage.
The maximum magnitude variations observed are $\sim0.2$\,mag, which are not larger than 3 times the typical measurement uncertainty. 
It is worth to note that part of the data were taken using different combinations of filters, cameras, and
telescopes which does not guarantee their homogeneity. Also, the faintness of the source in the optical
range prevents to obtain more accurate photometric measurements.

As in the optical range, our near-IR photometric measurements (performed in 2005 March and 2006 December) do not show 
significant variability with amplitude above $\sim0.3$\,mag. ($\sim 0.25$ or 30\%). 
A similar result is obtained when our near-IR measurements are compared with those of the 2-MASS survey in 1999 
(see Table~\ref{ta:phot}). In this case, maximum variability amplitudes of $\Delta J=0.3$\,mag and $\Delta
K_{\rm{S}}=0.4$\,mag are found, which contrasts to the factor of 2 flux density
increase reported by \citet{Terasranta05} at radio wavelengths for the same time range.
This suggests that the process responsible for the radio emission is not
connected to the one responsible for the optical and near IR emission. 

\input{table4}

\section{Discussion}
\label{sc:disc}

\subsection{Black hole mass and its accretion efficiency}

We use the empirical relationships provided by \citet{Vestergaard06} 
to estimate the central black hole mass in \object{NRAO~150}.
The black hole (BH) mass can be estimated from the 
luminosity and width of the broad component of the H$\beta$ line 
using expresion 6 in \citet{Vestergaard06}. 
Given the low S/N ratio of the H$\beta$ line detected in our 
spectral measurements, the use of the line luminosity is preferred against 
the value of the continuum at 5100\AA\ \citep[used in expression 5 of ][]{Vestergaard06}.
The measured broad component luminosity  is $5.6\pm1.2\times10^{43} erg\,s^{-1}$  
which increases up to $2.3\pm0.5 \times10^{44} erg\,s^{-1}$ after applying the
galactic extinction correction derived in this work (see Appendix~\ref{galextinction}).  
Combining these values with those for the line width (see Table~\ref{ta:fitlines}) 
the resulting black hole mass is $1.95\times 10^9 M_\odot$, which converts to $4.68\times10^9 M_\odot$ after
extinction correction. Using the estimate of the black hole mass we can also determine the
corresponding Eddington luminosity [$\mathrm{L}_{Edd} = 3.3\times 10^4 (M_{BH}/M_\odot) \, L_\odot$], which
results in a value of $1.54 \times 10^{14} \mathrm{L_\odot}.$ 
In order to compute the accretion rate, this value must be compared with the bolometric
luminosity of the accretion disk, which results in
$\mathrm{L_{disk}}\simeq 7.8\times 10^{12} \mathrm{L_\odot}$, and $\mathrm{L_{disk}}\simeq 46.9\times 10^{12} \mathrm{L_\odot}$
after extinction correction. 
The disk luminosity was computed by integrating the measured fluxes within the optical-UV 
range in the rest frame, this value is multiplied by $2\pi D^2$ to obtain the luminosity. 
Here it is assumed that the radio--mm and the X-ray emission
are related to the relativistic jet. This is justified given the uncoupled variability of the optical-UV spectral ranges
with regard to the radio--mm one as reported in Section~\ref{sec:varia}.
Thus the Eddington luminosity ratio is $\mathrm{L_{disk}/L_{Edd}} =0.30$. 

These values can be compared to those previously reported in the literature for similar objects.
The black hole mass of \object{NRAO~150} is above the highest masses found in the literature for low redshift 
AGNs \citep{Vestergaard06}, although the estimated value for \object{M~87}, is 
$3.4\times 10^9 M_\odot$ \citep{Graham07}, comparable to the one of \object{NRAO~150}.  
In contrast, the black hole mass of \object{NRAO~150} is well in the range of typical masses of 
luminous high redshift AGNs \citep{Shemmer04}. Several works
\citep[e.g.,][]{McLure04,Shen08} have tried to measure the dependence on redshift of the mean BH mass. 
They reach a common
result: the mean BH mass increases with redshift, although this 
dependence is dominated by the Malmquist bias \citep{Vestergaard08}.  
Recently, \citet{Labita09} claimed that the maximum BH mass evolves with z as $log(M_{BH(max)}/M_\odot) \sim 0.3\,z+9$, whereas the
maximum Eddington ratio (0.45) is essentially constant with z. However, \citet{Labita09} suggest that these results are
unaffected by the Malmquist bias. Our estimate for the BH mass in \object{NRAO~150} is slightly above the maximum value, and the Eddington 
ratio is also close to the maximum value,  which indicates that NRAO~150 possesses a very massive and efficiently accreting BH.

\subsection{FeII emission and weakness of the NLR emission features}
\label{sc:weakNLR}

As reported above, the spectrum of \object{NRAO~150} presents prominent H
recombination lines plus intense  FeII emission, in contrast with very weak or absent lines characteristic of the
NLR, such as [OIII] (see Fig.~\ref{fig:subFeII}). \citet{Netzer04} found that about one third of  
very high luminosity AGNs do not show strong [OIII] lines.  
In addition, an anticorrelation between EW(FeII)/EW(H$\beta$) and EW([OIII]) \citep{Boroson92,Yuan03,Netzer04} is reported.
In our \object{NRAO~150} spectra, the estimated EW for H$\beta$, FeII and [OIII] 
are around 55, 60 and 6, although the  uncertainties are large given the difficulty to deblend 
the spectral features. These values are consistent with the mentioned results. 
Using these values we checked that \object{NRAO~150} is placed in Fig. 9 of \citet{Netzer04} 
at the locus of high luminosity and high-redshift QSO sample. The location of \object{NRAO~150} in Fig. 3 of \citet{Netzer04} 
is intermediate between the one of PG QSOs \citep{Bennert02} and $z>2$ QSOs, according
to its H$\beta$ luminosity.  \citet{Netzer04} suggest that for 
highly luminous QSOs the NLR becomes extremely large (even larger than the size of the
largest galaxies) as expected from the "natural" $R_{NLR} \propto L_{ion}^{1/2}$
dependence, which makes NLR to disappear becoming dinamically unbound. 
The equivalent size of the NLR in \object{NRAO~150}, as predicted from the H$\beta$ luminosity 
($\sim 4\times10^{44} \mathrm{erg s^{-1}}$), is about 20~kpc \citep[Fig. 3 in][]{Netzer04}, which is well above the sizes 
commonly reported for typical NLR, hence explaining the weakness of the NLR emission features in our spectra.

\subsection{The spectral energy distribution}
\label{SED}

\begin{figure}
\includegraphics[width=9cm,angle=0]{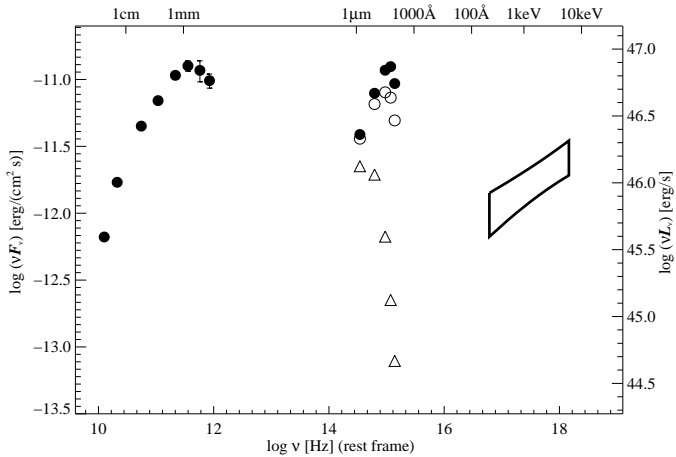}
\caption{\footnotesize{Spectral energy distribution (SED) of \object{NRAO~150} from the radio
to the X-ray spectral range after correction for galactic extinction.  Filled circles represent 
Galactic extinction estimated in this work (see Appendix~\ref{galextinction}), instead empty circles 
represent  extinction estimated from IRAS far-IR emission maps \citep{Schlegel98}.
Empty triangles indicate the extinction uncorrected
optical and near-IR measurements, note the dramatic change of the SED after extinction 
correction is applied.
The measurements presented here were obtained essentially contemporaneously during year
2005 (see text), except for the ROSAT X-ray data (bow tie), that were acquired between
August 1990 and February 1991.}\label{fi:sed}}
\end{figure}

Fig.~\ref{fi:sed} shows the spectral energy distribution (SED) of
\object{NRAO~150}. In addition to the photometric data presented here, we 
searched for data covering the widest possible wavelength range, from radio to X-rays. 
At radio bands, the lowest frequency measurements,
obtained at 4, 8.4, 22, and 43\,GHz with the VLA on March 2005, were taken from
the National Radio Astronomy Observatory\footnote{The National Radio Astronomy
Observatory is a facility of the National Science Foundation operated under
cooperative agreement by Associated Universities, Inc.} data archive. 
The next three measurements were observed at 86, 142, and 229\,GHz under the
general IRAM 30-meter Telescope AGN Monitoring Program \citep[][and references
therein]{Reuter97} on March 2005 (August 2005 for the larger frequency one).
The optical-UV rest frame observations corresponds to the near-IR and the 
optical observations presented here, and were acquired on March 2005 and November 2005, 
respectively.
On this spectral region, empty triangles indicate the extinction uncorrected
optical and near-IR measurements. Empty circles symbolize such measurements
corrected for the Galactic extinction estimated from IRAS far-IR emission maps
\citep{Schlegel98}, whereas filled circles indicate the same measurements
corrected for the more accurate Galactic extinction estimated here (see
Appendix~\ref{galextinction}).
The X-ray data, acquired by ROSAT from August 1990 and February 1991, 
were corrected by Galactic extinction as reported in \citet{Agudo07}. 
For the plot in Fig.~\ref{fi:sed}, a photon index $\Gamma=1.7\pm0.1$ was assumed
based on the synchrotron spectrum measured from the IRAM 30\,m observations
presented here at 86 and 142\,GHz (spectral index $\alpha=0.7\pm0.1$).

The SED of NRAO~150 shows two prominent bumps: one peaking around 1~mm and another
one at 2000~\AA. The  low frequency bump observed on the SED resembles those typical 
of high power flat-spectrum radio-loud sources 
\citep[e.g.,][]{GhiselliniTG09,Ghisellini09}. 
This first bump is attributed to strong
Doppler-bosted synchrotron relativistic-jet-emission.
Also the X-ray domain of the SED is typical from inverse Compton emission from
the jet, whereas the large luminosity  optical-UV bump
($L\approx10^{47}$\,erg/s at $\nu\approx10^{14.5}$\,Hz) is not so commonly reported for this kind of sources.
A prominent peak at optical-UV (rest frame) wavelengths is typical of
Seyfert galaxies and is thought to be produced by thermal emission from the AGN accretion disk.
However, there is an increasing number of high--power radio--loud blazars for which this
emission feature is being reported \citep[e.g,][]{Raiteri07,Abdo09,DAmmando09,GhiselliniT09,GhiselliniTG09,Ghisellini09}.
Indeed, nearly half of the radio loud blazars in the sample considered by
\citet{GhiselliniTG09} were recently reported to show a similar optical-UV large
excess. There is a consensus to consider this emission to come from thermal emission
from the accretion disk  \citep[e.g,][]{Raiteri07,Abdo09,DAmmando09,GhiselliniT09,GhiselliniTG09,Ghisellini09}, 
that is not outshined by the synchrotron bump from the jet when such bump peaks at $\nu<<10^{14.5}$\,Hz.
We also agree on such explanation for the case of \object{NRAO~150} for the
reasons outlined by \citet{Ghisellini09}.
Moreover, this also explains the above-reported lack of variability in the
observed optical and near-IR bands of \object{NRAO~150}, whereas the radio and mm
spectral ranges showed a factor 5 long term variability. This is
easily explained if such observed optical and near-IR (rest frame optical-UV)
emission is produced in the accretion disk and not in the relativistic jets
where the extreme variability comes from.



\subsection{Source classification}

From the VLBI point of view, \object{NRAO~150} shows a one-sided powerful
relativistic-superluminal jet typical of blazars, which indicates 
that the axis must be aligned close to the line of sight.
Moreover, its radio spectrum is typically flat, which classifies the source as 
a flat spectrum radio quasar (FSRQ). The optical rest frame spectrum shows
intense and broad ($\sim5000$\,km/s) H$_\alpha$ and H$_\beta$ emission, so the source 
is a type 1 AGN, consistent with the disk axis being aligned  close to the line of sight 
according to the standard AGN Unification scenario \citep{Urry95}. The optical and
radio properties are both consistent relative to the orientation of the radio jet and the
accretion disk.
The fact that the forbidden [OIII] lines are weak
points to a very luminous central engine, which is also consistent with the 
high Eddington rate determined from the optical/UV luminosity relative to the 
BH mass. The large accretion rate is consistent with the 
large radio loudness of \object{NRAO~150}. The BH mass is also among the highest derived for quasars 
\citep[][and references therein]{Vestergaard06,Vestergaard08,Ghisellini09}.

\section{Summary and conclusions}
\label{sc:conclu}

We have determined, for the first time, the cosmological distance of \object{NRAO~150}; one of the brightest radio 
to mm AGN sources in the northern sky; by means of its spectroscopic redshift (z=1.517 or $d_{\rm{L}}=11164$\,Mpc).
Given the low Galactic latitude of the source, its optical spectral lines have remained hidden to us for decades.
The new near-IR spectra presented here have revealed very intense H$\alpha$ emission together with 
less prominent H$\beta$  and FeII blended lines. 

The line width of H$\alpha$ points out that \object{NRAO~150}  is a broad line AGN, whose disk
axis must be oriented close to the line of sight, consistent with having a powerful one-sided 
superluminal radio--mm jet.
Based also on the radio to mm spectral and variability properties (including those observed with VLBI), we classify \object{NRAO~150} as a FSRQ blazar.
Consistent with previous observations of high redshift and highly luminous quasars, the optical rest frame spectrum 
 of the source show weak or absent spectral features typical of the NLR --such as [OIII]$\lambda 5007$~\AA.
This supports the idea that the highest luminosity accretion disks (and probably the most massive BH) favour the unbounding of their NLR 
by accretion disk radiation.   

Using empirical relationships between H$\beta$ line width and luminosity we 
estimate that the central engine in \object{NRAO~150} possesses a large 
black hole with mass $\sim5\times 10^{9} \mathrm{M_\odot}$. 
The radio to X--rays SED of the source shows two prominent bumps: one peaking at 
millimetre wavelengths --typical of synchrotron radiation from high power blazar jets--, and    
another in the near--UV ($\sim$2000\AA) 
that is attributed to thermal emission from the accretion disk.
The good spectral coverage of the disk emission allows us to make a reliable measurement of the 
bolometric luminosity of the disk, which turns out to be accreted at $\sim30$\,\% of the Eddington rate. 

This, together with the large BH mass of the source, its prominent BLR line luminosity, and its highly luminous synchrotron spectrum sets \object{NRAO~150} among the most powerful FSRQ blazars.
Such sources are also the most luminous hard X--ray and $\gamma$-ray inverse--Compton blazar emitters  \citep{GhiselliniT09,GhiselliniTG09,Ghisellini09}, and are routinely monitored by high energy space observatories as \emph{Fermi}.  
We predict that \object{NRAO~150} is one of such sources. However, it has not still been detected in $\gamma$-rays.  
The low Galactic latitude of the source implies a challenge for $\gamma$-rays observatories to detect it.
If that is possible in the future, modeling of the whole broad--band spectrum SED would allow for further investigation of the intrinsic 
physical parameters of this powerful blazar.

\begin{acknowledgements}
Financial support by the grant AYA2004-03136 from Plan Nacional de
Astronom{\ia}a y Astrof{\ia}sica is acknowledged.
 I. A. acknowledges support by an I3P contract with the Spanish ``Consejo
Superior de Investigaciones Cient\'{i}ficas", and by the Spanish ``Ministerio de
Ciencia e Innovaci\'on" and the European Fund for Regional Development through
grant AYA2007-67627-C03-03.
The Liverpool Telescope is operated on the island of 
La Palma by Liverpool John Moores University in the Spanish Observatorio del 
Roque de los Muchachos of the Instituto de Astrofisica de Canarias with financial 
support from the UK Science and Technology Facilities Council.
Some of the data published in this article were acquired with the IAC80 telescope 
operated by the Instituto de Astrof\'{i}sica de Canarias in the Observatorio del Teide.
We gratefully acknowledge H. Ungerechts for providing total flux                                                               
density mm measurements from the general IRAM 30-meter Telescope AGN Monitoring
Program. IRAM is supported by INSU/CNRS (France), MPG (Germany) and IGN
(Spain).

\end{acknowledgements}

\bibliographystyle{aa}
\bibliography{NRAO150_v7}{}

\begin{appendix}

\section{Galactic Extinction correction}
\label{galextinction}

We have already mentioned that the low galactic latitude of \object{NRAO~150}
and its
associated high galactic extinction has  defied the optical identification
and studies of this radio--source. In order to compute the intrinsic 
properties  of \object{NRAO~150} we have to correct for the galactic
extinction. At a first approach, 
we have taken the extinction from SIMBAD database, which is estimated 
from far infrared emission maps build by combining IRAS and COBE/DIRBE data \citep{Schlegel98}. 
The quoted value is E(B-V)=1.474, which implies $A_V=4.5$ to $A_K=0.54$. 

\begin{figure}
\includegraphics[width=8.7cm]{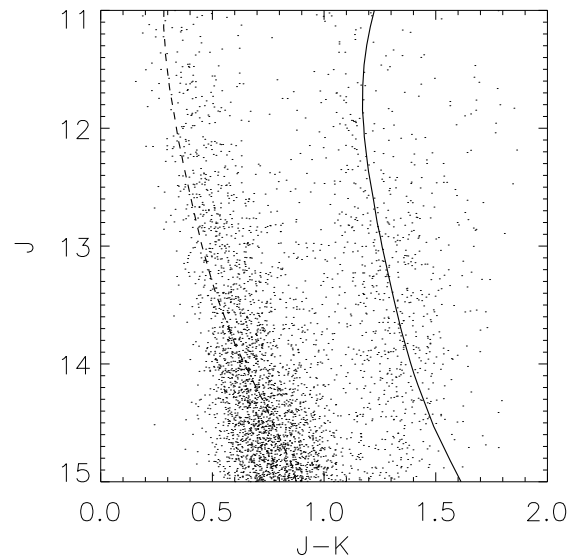}
\caption{\footnotesize{} Color-magnitude diagram of a field of $30 \times 
30$ arcmins, centered on \object{NRAO~150}. The two lines correspond to the 
theoretical Main Sequences of an isochrone with $\log t=9.1$ Gyr and 
z=0.019 of \citet{Girardi02}. Continous line fits the more distant 
and extincted stellar population while dashed line represents the fit
to a less extincted population of stars. \label{fig:cm}}
\end{figure}

In order to have another estimate of the galactic extinction 
towards the direction of \object{NRAO~150}, we built 
a color-magnitude diagram (Fig. \ref{fig:cm}) using the near infrared colors (J and 
K from 2MASS database \citealp{Cutri03}) of neighbouring stars.
The color magnitude ($J-K_s$,$J$) diagram stellar distribution 
was compared with the theoretical isochrones for 2MASS filters retrieved 
from  \citet{Girardi02}. We intend to recognize part of the Main Sequence 
in the color--magnitude diagrams. Indeed, we recognized two possible Main 
Sequences (see Fig. \ref{fig:cm}), one of them, more crowded, showing an excess 
$J-K\sim 0.7$ and a second sequence, less populated, with $J-K\sim 1.3$. 
We identified the first population to be in front of the Galactic disk and the other 
one to be behind the disk and presenting a higher galactic extinction.  

If we consider the solar metalicity, z=0.019, we found that the best 
possible fits for both stellar populations correspond to isochrones 
with $\log t=9.1$ Gyr (see figure \ref{fig:cm}). For these fits, we 
compute E(J-K)=$1.0 \pm 0.1$ for a stellar population located at 
distance module of 10.2, and E(J-K)$=0.12 \pm 0.3$ and 
a distance module of 9.5 for the less extincted and closer population. 
We are only interested in the largest value of E(J-K), which gives a
value E(B-V)=$1.7 \pm 0.2$ (E(B-V)=1.69~E(J-K)). This value is in reasonable  
agreement with the estimate provided by SIMBAD database. 

\end{appendix}

\end{document}

%% file: table1.tex
\begin{table}
\centering
\begin{tabular}{lcccc}
\multicolumn{5}{c}{SPECTROSCOPY} \\
\hline
\hline
Date  & Spec Range & Dispersion & T$_{exp}$  & Seeing \\
      &  \micron   & \AA/pix    & s     &   arcsec  \\
\hline
2005 March 23 & 0.89-1.33  & 6.1  & 3600 & $0\farcs7$ \\
2005 March 22 & 1.45-2.40  & 9.6 & 3600 & $0\farcs7$ \\
2007 Jan 2 & 0.89-1.33  & 6.1  & 3600 & $1\farcs2$ \\
2007 Jan 4 & 0.89-1.33  & 6.1  & 2400 & $1\farcs2$ \\
\hline                                                           
\end{tabular}
\begin{tabular}{lllcl}
\multicolumn{5}{}{} \\
\multicolumn{5}{c}{PHOTOMETRY} \\
\hline
\hline
Date & Instrum/Telesc & Filter & T$_{exp}$ (s) & Seeing \\
\hline
2005 March 22 & LIRIS/WHT & J  &  400 & $1\farcs4$ \\
              &           & K$_s$ &  400 & $1\farcs4$ \\
2005 Nov 2    & IAC--80   & V        & 9000 & $1\farcs5$ \\
	      &	       & R  	  & 6600 & $1\farcs3$ \\
	      &	       & I 	  & 6000 & $1\farcs3$ \\
2005 Nov 11   & IAC--80   & R        & 3600 & $1\farcs4$ \\
	      &	       & I  	  & 4800 & $1\farcs4$ \\
2006 Feb 03   & Liv. Tel. & V        &  600 & $1\farcs4$ \\
              & 	       & r'       &  300 & $1\farcs2$ \\
              & 	       & i'       &  200 & $1\farcs2$ \\
2006 Feb 17   & Liv. Tel. & r'       & 1800 & $1\farcs1$ \\
              & 	       & i'       &  900 & $1\farcs0$ \\
2006 Feb 20   & Liv. Tel. & r'       & 1800 & $1\farcs3$ \\
              & 	       & i'       &  900 & $1\farcs2$ \\
2006 Feb 22   & Liv. Tel. & r'       & 1400 & $0\farcs9$ \\
2006 Aug 22   & Liv. Tel. & r'       & 1400 & $0\farcs9$ \\
2006 Sep 18   & Liv. Tel. & r'       & 1400 & $0\farcs9$ \\
2006 Sep 30   & Liv. Tel. & r'       & 1400 & $0\farcs9$ \\
2006 Oct 21   & Liv. Tel. & r'       & 1400 & $0\farcs9$ \\
2006 Nov 13   & Liv. Tel. & r'       & 1400 & $0\farcs9$ \\
2006 Nov 22   & Liv. Tel. & r'       & 1400 & $0\farcs9$ \\
2006 Dec30    & LIRIS/WHT & J,       &  300 & $1\farcs1$ \\
              & 	       & K$_s$    &  250 & $1\farcs1$ \\
2007 Jan 8   & Liv. Tel. & r'       & 1400 & $0\farcs9$ \\
2007 Jan 14   & Liv. Tel. & r'       & 1400 & $0\farcs9$ \\
2007 Jan 21   & Liv. Tel. & r'       & 1400 & $0\farcs9$ \\
\hline                                                           
\end{tabular}
\caption[]{Observing log}
\label{ta:obslog}
\end{table}

%% file: table2.tex
\begin{table}
\centering
\begin{tabular}{llcc}
\multicolumn{4}{c}{Near IR} \\
\hline
\hline
Date & JD  & J & Ks  \\
\hline
1999 Oct 13 & 51465 & 16.49$\pm$0.12$^*$  & 14.37$\pm$0.08$^*$ \\
2005 Mar 23 & 53453 & 16.78$\pm$0.06  & 14.81$\pm$0.05 \\
2006 Dic 30 & 54100 & 16.52$\pm$0.04  & 14.54$\pm$0.05  \\
\hline                                                           
\end{tabular}
\begin{tabular}{llccc}
\multicolumn{5}{}{} \\
\multicolumn{5}{c}{Optical} \\
\hline
\hline
Date & JD & V & $r'$  &  $i'$ \\
\hline
2005 Nov 2 & 53677.45  & 22.9$\pm$0.9 & 20.67$\pm$0.08 & 19.43$\pm$0.10 \\
2005 Nov 11 & 53683.4  & ...          & 20.81$\pm$0.07 & 19.72$\pm$0.04 \\
2006 Feb 3 & 53770.35  & ...          & 20.84$\pm$0.04 & 19.19$\pm$0.02 \\
2006 Feb 17 & 53784.35 & ...          & 20.58$\pm$0.03 & 19.32$\pm$0.05 \\
2006 Feb 20 & 53787.48 & ...          & 20.60$\pm$0.06 & 19.68$\pm$0.05 \\
2006 Feb 22 & 53789.38 & ...          & 20.70$\pm$0.05 & 19.57$\pm$0.06 \\
2006 Aug 22 & 53970.65 & ...          & 20.72$\pm$0.10 & ... \\
2006 Sep 18 & 53997.51 & ...          & 20.88$\pm$0.02 & ... \\
2006 Sep 30 & 54009.56 & ...          & 20.60$\pm$0.05 & ... \\
2006 Oct 21 & 54030.46 & ...          & 20.62$\pm$0.01 & ... \\
2006 Nov 13 & 54053.39 & ...          & 20.68$\pm$0.02 & ... \\
2006 Nov 22 & 54062.44 & ...          & 20.78$\pm$0.02 & ... \\
2007 Jan 8 & 54109.38 & ...           & 20.85$\pm$0.01 & ... \\
2007 Jan 14 & 54115.35 & ...          & 20.67$\pm$0.04 & ... \\
2007 Jan 21 & 54122.40 & ...          & 20.81$\pm$0.09 & ... \\
\hline                                                           
\end{tabular}
\caption[]{NRAO 150 - Optical near-IR Photometry. \\
$^*$ Values corresponding to 2-MASS PSC. }
\label{ta:phot}
\end{table}

%% file: table3.tex
\begin{table*}
\begin{tabular}{lccccc}
\hline
\hline
ID & 2MASS--ID & R.A. & Dec (J2000)          & $r'$  & $i'$ \\
   &           & (03:59:ss) & (+50:$'$:$''$) &       &       \\
\hline
1 & J03593205+5057513 & 32.05 & 57:51.4 & $20.12\pm 0.04$ & $19.10 \pm 0.07$ \\
2 & J03593386+5057522 & 33.87 & 57:52.2 & $20.12\pm 0.02$ & $18.48 \pm 0.01$ \\
3 &    ...            & 31.86 & 57:45.0 & $20.86\pm 0.10$ & $19.79 \pm 0.14$ \\
4 & J03592824+5058196 & 28.24 & 58:19.6 & $19.62\pm 0.01$ & $18.66 \pm 0.07$ \\
5 &    ...            & 56.58 & 57:47.3 & $21.00\pm 0.07$ & $19.75 \pm 0.14$ \\
6 & J03592625+5057054 & 26.26 & 57:05.1 & $18.98\pm 0.01$ & $18.04 \pm 0.09$ \\
7 & J03592814+5058277 & 28.18 & 58:27.7 & $18.76\pm 0.01$ & $17.70 \pm 0.04$ \\
\hline                                                           
\end{tabular}
\caption[]{Photometry of reference stars near NRAO~150. \\
This table could be published only in the electronic version.\label{tab:comp}}
\end{table*}

%% file: table4.tex
\begin{table*}
\begin{tabular}{lccccccc}
\hline \hline
             & Center & \multicolumn{2}{c}{FWHM}    & \multicolumn{2}{c}{Flux}  & \multicolumn{2}{c}{Lum.}\\
Line         & (\AA)  & (\AA)  & km/s               & Obs.     & Ext. Cor.  & Obs.     & Ext. Cor. \\
\hline 
H$\alpha$(n) - 2005 March & 16501.1$\pm$0.7 & 72$\pm$1.7 &  1458 & 22.0$\pm$0.5 & 51.7$\pm$1.2 & 32.9$\pm$0.7 & 77.1$\pm$1.8 \\ 
H$\alpha$(b) - 2005 March & 16439$\pm$7     & 283$\pm$16 &  5745 & 21.7$\pm$0.7 & 51.0$\pm$1.6 & 32.4$\pm$1.0 & 76.1$\pm$2.4  \\ 
H$\beta$(n)  - 2005 March & 12255$\pm$5     & 54$\pm$1.2 &  1458 & 2.97$\pm$0.3 & 12.0$\pm$1.2 & 4.4$\pm$0.4 & 17.9$\pm$0.6  \\ 
H$\beta$(b)  - 2005 March & 12209$\pm$5     & 210$\pm$12 &  5745 & 3.74$\pm$0.8 & 15.1$\pm$3.2 & 5.6$\pm$1.2 & 23$\pm$5  \\ 
H$\beta$(n)  - 2007 Jan   & 12254$\pm$9     & 54$\pm$1.2 &  1458 & 2.14$\pm$0.4 &  8.6$\pm$1.6 & 3.2$\pm$0.6 & 13$\pm$2  \\ 
H$\beta$(b)  - 2007 Jan  & 12208$\pm$9     & 210$\pm$12 &  5745 & 3.45$\pm$0.9 & 13.9$\pm$3.6 & 5.1$\pm$1.3 & 21$\pm$5  \\
\hline
\end{tabular}
\caption[]{Flux units are $10^{-15} \mathrm{erg} \mathrm{s}^{-1} \mathrm{cm}^{-2}$. 
Luminosity units are $10^{43} \mathrm{erg} \mathrm{s}^{-1} $\label{ta:fitlines}}
\end{table*}
